\documentclass[doublecol]{epl2} 

\usepackage{graphics}
\usepackage{epsfig}
\usepackage{amssymb}
\usepackage{subfigure}
\usepackage{pifont}

\title{Mechanical response of an inclined frictional granular layer approaching unjamming}
\shorttitle{Mechanical response of an inclined frictional granular layer} 

\author{A.P.F. Atman\inst{1} \and P. Claudin\inst{2} \and G. Combe\inst{3} \and R. Mari\inst{2}}
\shortauthor{A.P.F. Atman \etal}

\institute{                    
  \inst{1} Departamento de F\'{\i}sica e Matem\'atica and National Institute of Science and Technology for Complex Systems, Centro Federal de Educa\c c\~ao Tecnol\'ogica de Minas Gerais, CEFET--MG,
Av. Amazonas 7675, 30510-000, Belo Horizonte-MG, Brazil.\\
  \inst{2} Laboratoire de Physique et M\'ecanique des Milieux H\'et\'erog\`enes, (PMMH UMR 7636 CNRS -- ESPCI -- Univ. P. et M. Curie -- Univ. Paris Diderot), 10 rue Vauquelin, 75231 Paris Cedex 05, France.\\
  \inst{3} UJF-Grenoble 1, Grenoble-INP, CNRS UMR 5521, 3SR Lab., B.P. 53, 38041 Grenoble Cedex 09, France.
}

\abstract{
We present an orthotropic elastic analysis of frictional granular layers under gravity by studying their stress response to a localized overload at the layer surface for several substrate tilt angles. The distance to the unjamming transition is controlled by the tilt angle $\alpha$ with respect to the critical angle $\alpha_c$. We find that the shear modulus of the system decreases with $\alpha$, but reaches a finite value as $\alpha \to \alpha_c$. We also analyze the vibration modes of the system and show that the soft modes play an increasing, though not crucial, role approaching the transition.}

\begin{document}

\maketitle

Various amorphous materials, and granular media in particular, exhibit a so-called jamming transition between rigid and flowing states. The nature of this transition has been investigated during the last decade, see recent reviews \cite{vH10, LNvSW10}.  Most granular studies have focused on frictionless discs or spheres, typically controlled in volume fraction $\phi$ or in pressure $P$ \cite{oHLLN02, oHSLN03, MSLB07}, showing that the jamming transition is critical (scaling exponents, diverging length scale) \cite{oHLLN02, WNW05, EvHvS09} and related to isostaticity \cite{R00, TW99, M01, oHLLN02, AR07}. As the system looses its mechanical rigidity at the transition, its shear modulus $G$ is found to vanish as a power law with respect to the distance to jamming $\phi-\phi_c$, where $\phi_c$ is the critical volume fraction. Vibrational mode analysis have been performed on these frictionless granular systems \cite{oHLLN02, SLN05, DHRR05, WNW05, SBoHS11}, or similarly on soft glassy materials \cite{GCSKB10, KGMI10, CEZCYHBDvSLY10}, reporting an increasing number of soft modes, corresponding to collective low-energy motion of the particles, as the transition is approached.

\begin{figure*}[t!]
\centerline{\includegraphics[width=0.95\textwidth]{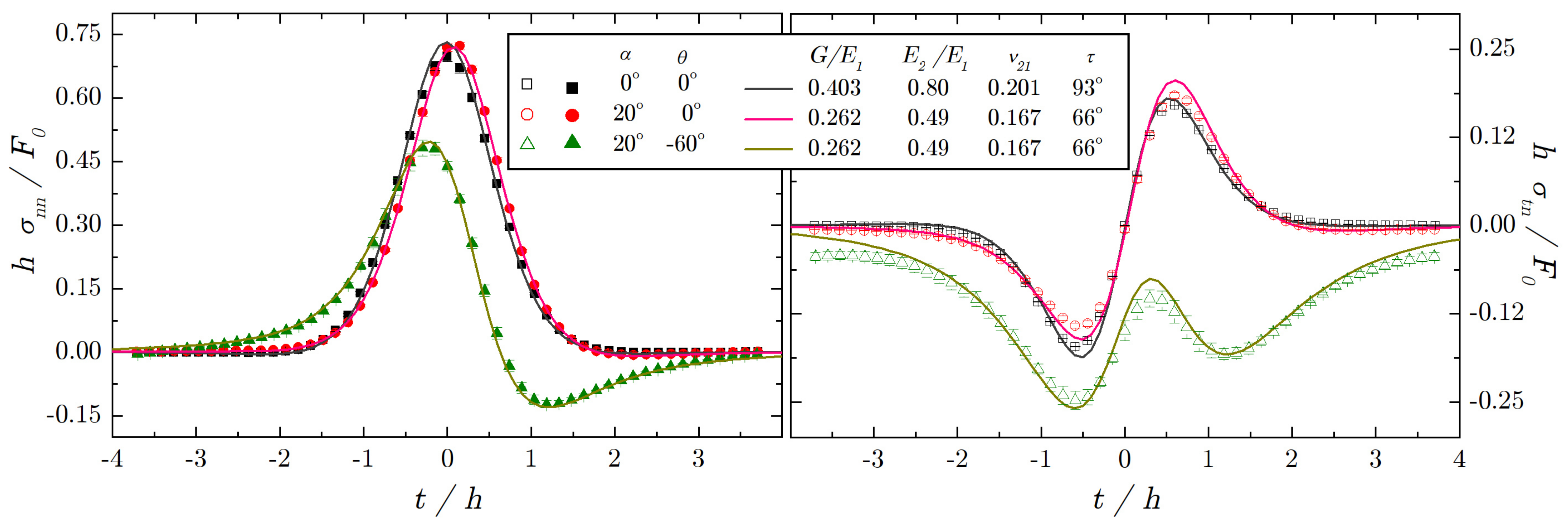}}
\caption{(a) Normal stress response profile $\sigma_{nn}$ for two values of $\alpha$, and two values of $\theta$ (see legend). Symbols: numerical data from the MD simulations (GG preparation, \cite{ABGRCCBC05}). Solid lines: best elastic fit (see parameters in legend). The stresses are normalized by $F_0/h$ and the distance by $h$. (b) Same for the shear stress response $\sigma_{tn}$.}
\label{fig:StressResponseProfiles}
\end{figure*}

\begin{figure}[t!]
\centerline{\includegraphics[width=0.45\textwidth]{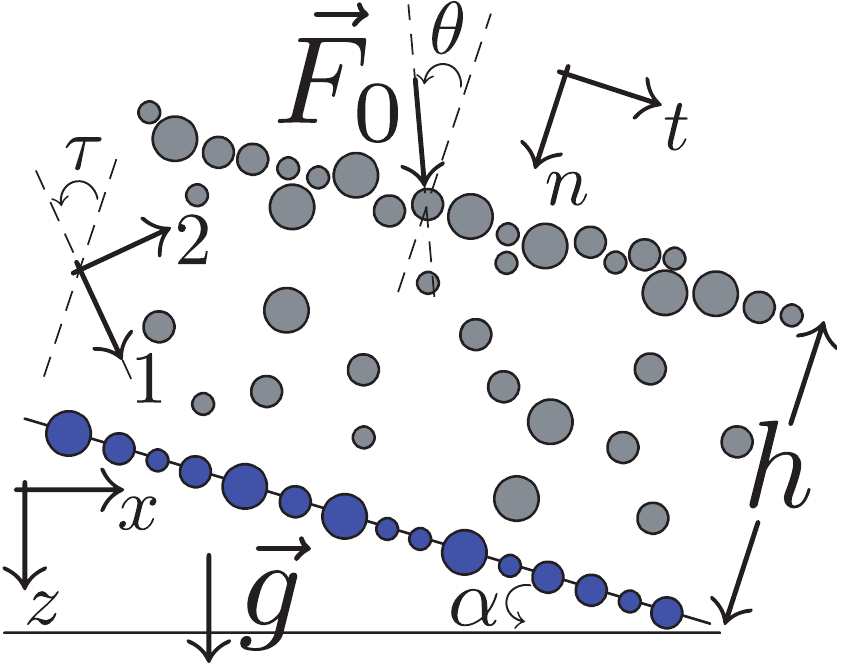}}
\caption{Schematics of the system and notations.  $x$ is the horizontal axis. $z$ is the vertical one, along which acts gravity $\vec g$. The granular layer, of average thickness $h$, is inclined at an angle $\alpha$ with respect to horizontal. $t$ and $n$ are the axis respectively tangential and normal to the layer.  A localized force ${\vec F}_0$, which makes an angle $\theta$ with respect to $n$, is applied on a grain close to the surface of the layer. The stress responses $\sigma_{nn}$ and $\sigma_{tn}$ to this overload are measured at the bottom of the layer (fixed grains in blue). Axis $(1,2)$, making an angle $\tau$ with respect to $(n,t)$, are those of the orthotropic elastic analysis.}
\label{fig:Scheme}
\end{figure}

Much fewer studies have dealt with frictional grains in this context and they have mostly considered homogeneous systems under isotropic pressure \cite{AR07, SvHESvS07, HvHvS10, HSvSvH10, S10, BDBB10}. In the frictional case, the Liu-Nagel jamming concept \cite{LN98,LN10} must be revised \cite{BZCB11}. In particular, jamming and isostatic points do not coincide any more \cite{vH10}, and one thus can expect a finite shear modulus at the transition. In this letter, we consider static layers of frictional grains under gravity, by means of two-dimensional discrete element simulations (standard Molecular Dynamics \cite{DEM}), and investigate their mechanical properties through the analysis of their stress response to a localized overload $\vec F_0$ at the layer surface. The layers are prepared at a fixed angle $\alpha$ with respect to the horizontal (see Fig.~\ref{fig:Scheme} for notations), and unjamming is approached as $\alpha$ is close to $\alpha_c$, the critical value above which static layers cannot be equilibrated at that angle and always flow. Note that this unjamming point $\alpha_c$ is close in spirit to the situation of a jammed solid sheared up to its yield-stress~\cite{HB09}. It is also close, but different, to progressively tilted granular layers, which eventually loose their mechanical stability, see e.g., \cite{SVR02, HBDvS10}. In our simulations, the volume fraction in the layer is fairly uniform all through the layer depth. Its value depends very weakly on the inclination angle, varying from $\phi \simeq 0.823$ at $\alpha=0$ to $\phi \simeq 0.816$ at $\alpha_c$ (these are for the GG preparation, see below). These values are always larger -- though not much -- than the critical value, estimated in our system at $\phi_c \simeq 0.815$ \cite{dCEPRC05,S10, OH11}. Then, here, the control parameter for the jamming/unjamming transition is the sole angle $\alpha$. This situation is therefore qualitatively different to the homogeneous configurations submitted to isotropic pressure cited above, and is effectively closer to an experimental set-up.

The numerical model is that described in \cite{ABGRCCBC05, GACCG06}, with $N=3600$ polydisperse frictional discs coupled, when overlapping, by normal and tangential linear springs, tangential forces being limited by the Coulomb condition with a friction coefficient $\mu=0.5$. Two system preparations have been considered: a grain-by-grain (GG) and a rain-like (RL) deposition of the particles on a rough substrate consisting of fixed but size-distributed particles, inclined at the desired angle $\alpha$. The layer is prepared when all grains have reached static equilibrium (see \cite{ABGRCCBC05} for a description of these protocols). No external pressure applied to the topmost layer of particles, i.e. the pressure in the system is due solely to the gravitational force acting on the particles themselves. The typical thickness of the layer is $\simeq 23$ grain diameters, i.e. a system aspect ratio around $1/7$. Above a certain inclination $\alpha_c$, the simulations preparing the packing before the response procedure do not converge towards a static layer -- the grains always flow. The `solid-liquid' transition is rather abrupt ($\Delta \alpha \simeq 0.5^\circ$), allowing us for a well defined value of this critical angle: $\alpha_c \simeq 20.75^\circ$ for the GG and $\alpha_c \simeq 20^\circ$ for the RL preparations respectively. For the sake of concision, except otherwise stated, the displayed data are those obtained with the GG preparation.

Experimental and numerical works have shown that the linear stress response of granular systems to a point force is well described by (possibly anisotropic) elasticity \cite{SRCCL01, ABGRCCBC05, GG02, LTWB04, GG05, ABGRCBC05}. Here we show that an orthotropic elastic model is able to reproduce quantitatively the normal and tangential stress bottom profiles, and that the corresponding shear modulus $G$ decreases with $\alpha$, in parallel to a decrease of the coordination number and minor changes in the micro-structure (fabric). However, it does not vanish at the transition, but reaches a finite value as $\alpha \to \alpha_c$. We finally analyze the vibration modes of the system and show that the soft modes play an increasing, though not crucial, role approaching the transition.

\section{Stress response functions}
Once a layer is deposited, stabilized in an equilibrium state, an additional force $\vec F_0$ is applied on a grain close to the free surface, and a new equilibrium state is reached (see \cite{ABGRCCBC05} for details). Taking the difference between the states after and before the overload, one can compute the contact forces in response to $\vec F_0$. Introducing a coarse graining length $w$, the corresponding stress response can be determined. Taking $w$ of the order of few mean grain diameters (here $w =  6 \left < d \right >$) as well as an ensemble averaging of the data (here $N_e \sim 150$ realisations for each tilt angle $\alpha$), make the stress profiles quantitatively comparable to a continuum theory \cite{GACCG06}, such as elasticity, as discussed below. The amplitude of the overload was kept constant for all simulations: $F_0 =  1.0 \left < m \right > g$, where $\left < m \right >$ is the average mass of the grains. This value is  sufficiently small to ensure a linear \cite{ACCGG09, ACC09} and reversible response of the system for all values of $\alpha$, including close to $\alpha_c$.

Three examples of stress bottom profiles $\sigma_{nn}(t)$ and $\sigma_{tn}(t)$ for the GG preparation are displayed in Fig.~\ref{fig:StressResponseProfiles} for two values of the slope $\alpha$ of the layer and two values of the angle $\theta$ that the overload force makes with the normal direction (see Fig.~\ref{fig:Scheme}). The stress profiles computed for the RL preparation have approximatively the same qualitative behavior.

\begin{figure}[t!]
\centerline{\includegraphics[width=0.45\textwidth]{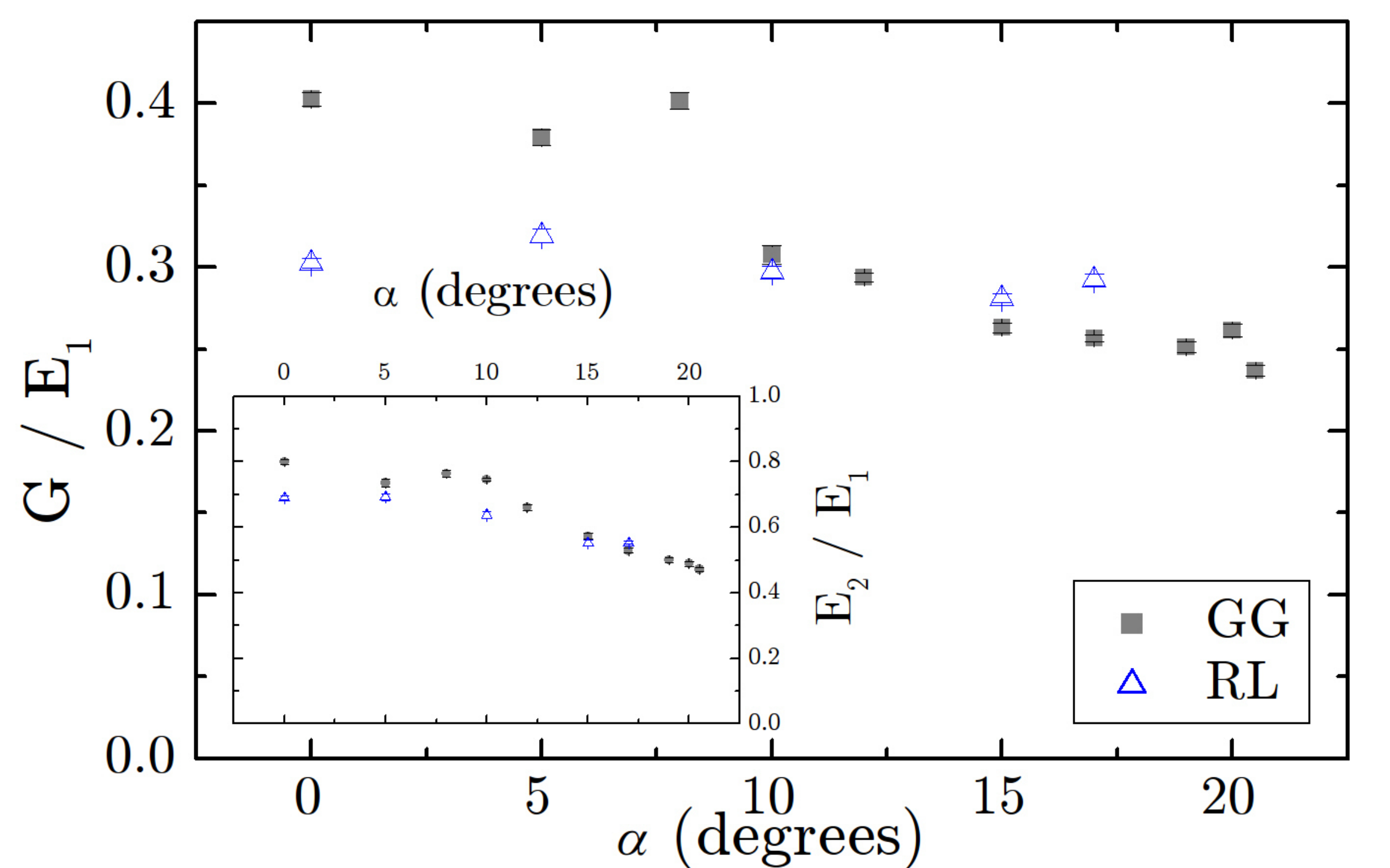}}
\caption{Modulus ratio $G/E_1$ determined from the elastic fit, as a function of $\alpha$, for two different layer preparations (see legend). Inset: Same for $E_2/E_1$.}
\label{fig:Gofvarphi}
\end{figure}

\section{Orthotropic elastic analysis}
In order to reproduce the stress response profile, isotropic elasticity is not enough \cite{ABGRCCBC05}, and we consider here orthotropic elasticity, characterized by a stiff axis ($1$) and a soft one ($2$) associated to two Young moduli $E_1$ and $E_2<E_1$. Two Poisson coefficients $\nu_{12}$ and $\nu_{21}$ must also be introduced. However, they are not independent and verify $\nu_{12}/E_1 = \nu_{21}/E_2$. Finally, $G$ is the shear modulus, so that the stress-strain relation, in the orthotropic axis writes:
\begin{equation}
\left (
\begin{tabular}{c}
$\epsilon_{11}$\\
$\epsilon_{22}$\\
$\epsilon_{12}$
\end{tabular}
\right )
=
\left (
\begin{tabular}{ccc}
$\frac{1}{E_1}$			& $-\frac{\nu_{21}}{E_2}$		& $0$\\
$-\frac{\nu_{12}}{E_1}$	& $\frac{1}{E_2}$			& $0$\\
$0$					& $0$					& $\frac{1}{2G}$
\end{tabular}
\right )
\left (
\begin{tabular}{c}
$\sigma_{11}$\\
$\sigma_{22}$\\
$\sigma_{12}$
\end{tabular}
\right ).
\end{equation}
Elastic energy is well defined if all moduli $E_1,E_2,G$ are positive and $1-\nu_{12}\nu_{21}>0$. A last parameter of this modeling is the angle $\tau$ that the axis $(1,2)$ make with $(n,t)$ (see Fig.~\ref{fig:Scheme}).

The stress responses have been analytically computed in \cite{OBCS03} for a semi-infinite layer ($h \to \infty$). For a finite layer thickness, a numerical integration, in the spirit of \cite{SRCCL01}, must be done. Rough bottom boundary conditions (zero displacement) are imposed. We can adjust four dimensionless parameters (two modulus ratios $G/E_1$ and $E_2/E_1$, a Poisson coefficient $\nu_{21}$ and the orthotropic angle $\tau$) to reproduce, for a given $\alpha$, the numerical data for all $\theta$. As shown in  Fig.~\ref{fig:StressResponseProfiles} (solid lines), the fitting functions quantitatively describe the numerical data.

In Fig.~\ref{fig:Gofvarphi}, we show the behavior of the elastic
modulus ratios $G/E_1$ and $E_2/E_1$ as the slope $\alpha$ is
larger. They both decrease, almost by a factor of two, from a fairly
constant value at small $\alpha$ to a lower one close to
$\alpha_c$. In particular, $G$ does not vanish close to the critical
angle, in agreement with the observation that frictional granular
systems remain hyperstatic at the unjamming transition
\cite{AR07,SvHESvS07,HvHvS10}. Such a discontinuous behaviour has been
seen in simulations by Otsuki and Hayakawa \cite{OH11} investigating
the rheology of sheared frictional grains close to jamming, and in
experimentally created shear-jammed states reported in
\cite{BZCB11}. The ratio $G/E_2$ is almost constant (not shown). This
behavior is also found for the RL preparation, although the variations
of $G/E_1$ are less pronounced. This softening is finally consistent with acoustic experiments on a granular packing in the vicinity of the transition \cite{BAC08}.

\begin{figure}[t!]
\centerline{\includegraphics[width=0.45\textwidth]{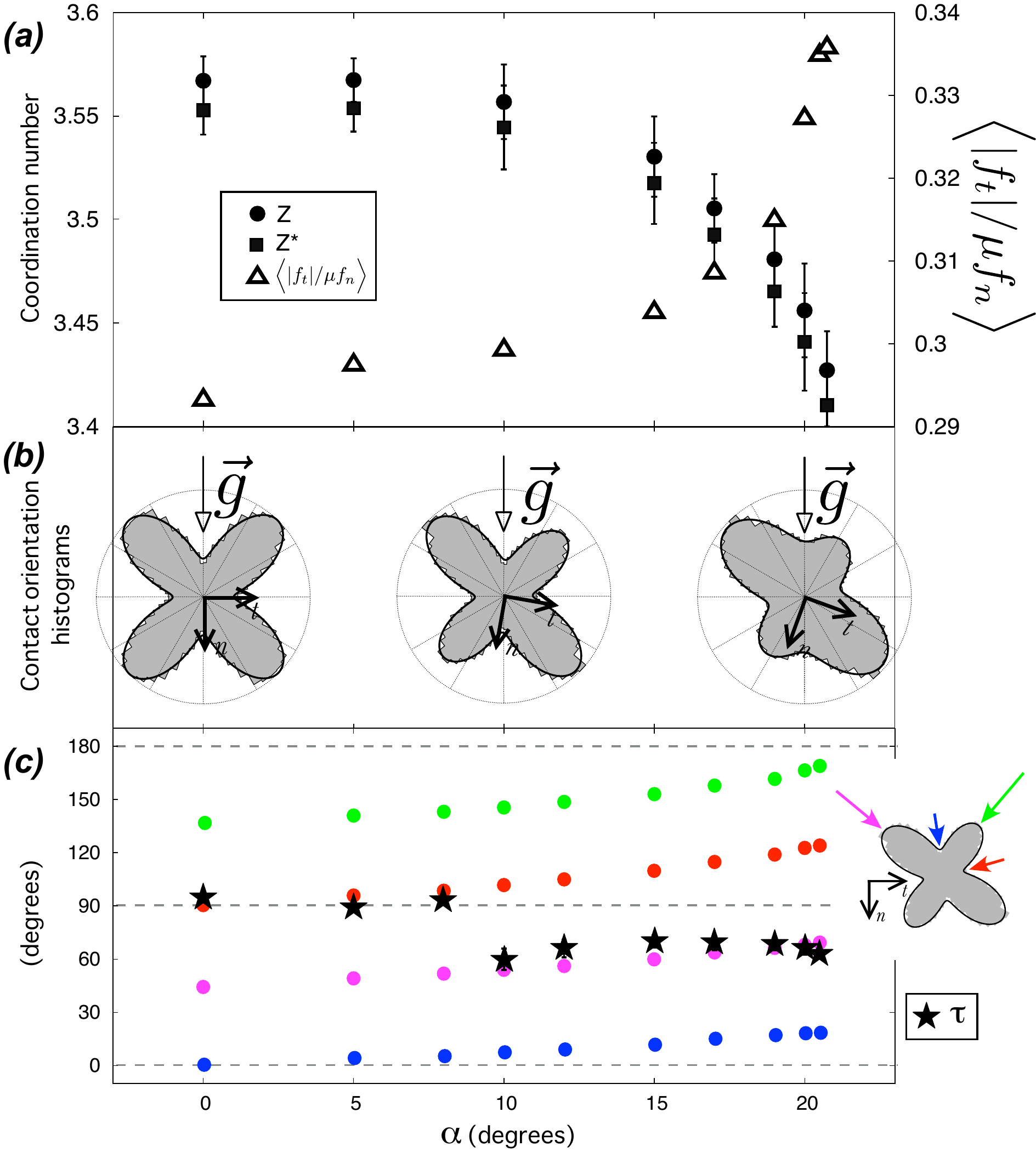}}
\caption{(a) Evolution of the coordination number $Z$ (\ding{108}) with the inclination of the layer $\alpha$. Taking into account `rattlers' (see text) gives a modified coordination number $Z^*$ ($\blacksquare$). Right $y$-axis: relative importance of friction mobilisation at contact ($\bigtriangleup$). (b) Contact angle polar distributions (GG preparation) at $\alpha=0, 10, 20^\circ$. Solid black line: fourth-order Fourier fit. Gravity is vertical (black arrow). (c) Fitted orthotropic elastic angle $\tau$ as a function of $\alpha$ ($\bigstar$). The four characteristic angles of the contact angle distribution, computed with respect to the direction $n$, are also shown -- these angles corresponds to the directions of the lobes, and those in between the lobes, see sketch and corresponding coloured arrows.}
\label{fig:Micro}
\end{figure}

\section{Microscopic variables}
In addition to the above global mechanical properties of the system, we have studied the evolution of various microscopic quantities with $\alpha$. The first one of interest is the coordination number $Z$, i.e. the average number of contacts per grain, here computed in the bulk of the layer, where it is fairly uniform -- it obviously drops down close to the surface. $Z$ monotonously decreases with $\alpha$ (Fig.~\ref{fig:Micro}a) and stays always far from the isostatic value $Z_{\rm iso}=3$ (for frictional grains in 2D). As evidenced by the comparison of the curves in figures~\ref{fig:Gofvarphi} and \ref{fig:Micro}a, the modulus ratio $G/E_1$ is not found to be a linear function of $Z-Z_{\rm iso}$, in contrast with the finding of \cite{SvHESvS07} on homogeneous frictional systems, close to isostaticity. Grains of the bulk that only carry their own weight do not contribute much to the global stability of the contact network. As for so-called rattlers in gravity-free packings (see \cite{bookDEM}, chap. 6), these grains can be removed from the contact counting, leading to a modified coordination number of the layer $Z^*$ (see Fig.~\ref{fig:Micro}a). However, we have found that their number is roughly independent of $\alpha$, so that the relation between $G/E_1$ and $Z^*$ is not linear either. We have also studied the friction mobilisation at the contact level. In our MD simulations, the number of contacts with a ratio of the tangential force $f_t$ to the normal force $f_n$ strictly equal to the microscopic friction $\mu$ is zero when static equilibrium is reached. However, some of them are effectively close to the Coulomb criterion, and we have computed the average $\langle \frac{|f_t|}{\mu f_n} \rangle$. This quantity, displayed in Fig.~\ref{fig:Micro}a, increases as $\alpha \to \alpha_c$, but its overall variation is weak (see right $y$-scale).

Finally, we have studied the contact angle distribution. Three of these distributions are represented as polar diagrams for $\alpha=0$, $10$ and $20$ degrees in Fig.~\ref{fig:Micro}b. The four strongly pronounced lobes are typical of the GG preparation \cite{bookDEM} (chap. 6). The vertical and horizontal directions are always in between these lobes. When the layer is horizontal, the orthotropic stiff and soft directions are also found to be (almost) along the horizontal and vertical axis respectively. Note that the fitting procedure effectively gives here $\tau=93^\circ$ in this case, while $\tau=90^\circ$ (or $0^\circ$) would have been expected for symmetry reasons. This indicates the typical precision we have on the measure of this orthotropic angle. Close to the critical slope, however, the orthotropic orientations are close to those of the lobes, the stiff one being in the direction of the slope. As evidenced in Fig.\ref{fig:Micro}c, the transition between these two microscopic configurations occurs around $\alpha \simeq 9^\circ$, i.e. well below $\alpha_c$, in correspondence with the drop by a factor of $2$ of $G/E_1$ between $8$ and $10^\circ$ (see Fig.~\ref{fig:Gofvarphi}).

\begin{figure}[t!]
\centerline{\includegraphics[width=0.45\textwidth]{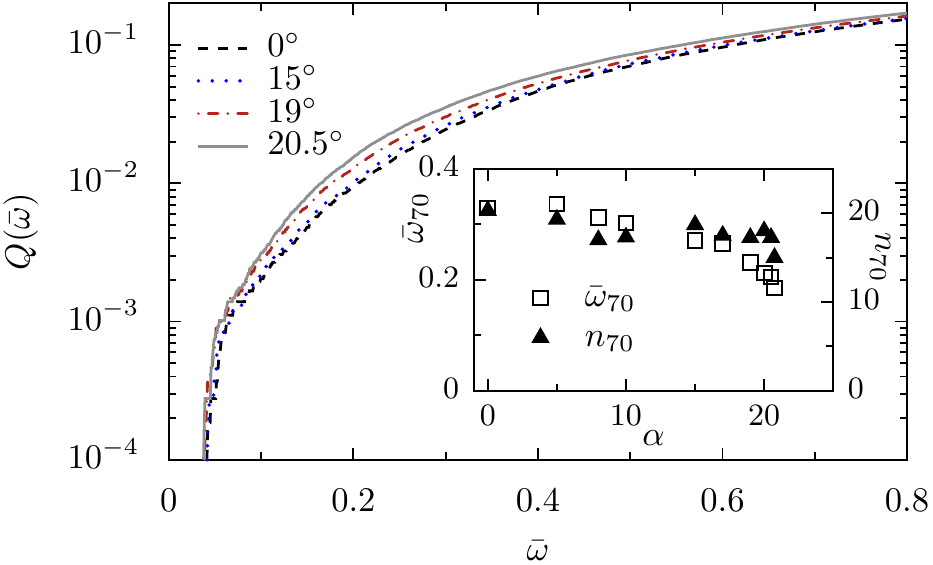}}
\caption{Cumulative density of vibration modes $Q(\bar \omega)$ for four different values of $\alpha$ (see legend). Inset: Number of modes $n_{70}$, which are necessary to represent $70 \%$ of the displacement response, and corresponding highest frequency $\bar \omega_{70}$, as a function of $\alpha$.}
\label{fig:Modes}
\end{figure}

\section{Vibration mode analysis}
For each tilt angle $\alpha$, we have computed the $3N$ vibration modes of the layer, performing a harmonic approximation of the energy around the equilibrium point reached by deposition of the grains. We only studied layers obtained by GG deposition. They do not show contacts being at the Coulomb threshold. This avoids the issue of the relevance of the harmonic approximation for such contacts~\cite{HvHvS10}. Each contact between two grains is thus being considered as two linear springs (normal and tangential stiffnesses $k_n$ and $k_t$; here $k_t=0.75 k_n$). The reference vibration frequency is that of a single spring-mass system $\omega_0 = \sqrt{k_n/\left < m \right >}$. We denote $D(\bar \omega)=(3N)^{-1}\sum_{{\rm mode}\, i} \delta(\bar \omega- \bar \omega_i)$ and $Q(\bar \omega) = \int_0^{\bar \omega} D(\bar \omega')\, {\rm d}\bar \omega'$, the density and cumulative density of states, where $\bar \omega = \omega/\omega_0$ is the normalized frequency.

The function $Q$ is displayed in Fig.~\ref{fig:Modes} for four values of the slope $\alpha$. No excess of vanishing frequencies are observed as the transition is approached, which is expected for a system that stays hyperstatic \cite{WSNW05}. There is no single zero frequency mode appearing at the transition either. The criticality of the layer thus cannot be described by the emergence of an energetically costless direction in configuration space. However, some rather soft modes ($0.1 \lesssim \bar \omega \lesssim 0.5$) are enhanced when $\alpha \to \alpha_c$. We have also computed the eigenfunctions corresponding to these modes. The displacement field in response to the overload with $\vec F_0$ can be decomposed on these functions, as they form a mathematical basis. We denote $c_i$ the coefficients of this decomposition. A large part of the decomposition is owned by the $~20$ lowest frequency modes. Hence, we can compute the number of modes $n_{70}$ which are necessary to represent $70 \%$ of the displacement field, and the highest frequency $\bar \omega_{70}$ of these $n_{70}$ modes, by looking for the lowest frequency $\bar \omega_{70}$ such that $\sum_{i | \bar \omega_i < \bar \omega_{70}} c_i^2 > 0.70 $. Both $n_{70}$ and $\bar \omega_{70}$ are plotted against $\alpha$ in the
inset of Fig.~\ref{fig:Modes}. While $n_{70}$ stays approximatively constant and around $20$, $\bar \omega_{70}$ is found to decrease from $\simeq 0.3$ by a factor of two when $\alpha$ varies from zero to $\alpha_c$, without however showing a clear singularity at the transition. This means that the response is progressively better represented by softer modes, even if these modes do not trigger the critical behavior of the layer close to the transition.

\section{Conclusions}
To sum up, we have simulated 2D frictional and polydisperse granular layers under gravity inclined at an angle $\alpha$, and investigated their mechanical and microscopic properties when the unjamming transition is approached. This work tells us what to expect in real experiments, i.e. a layer that becomes elastically softer as $\alpha \to \alpha_c$, but not to the point at which the system would loose its rigidity before avalanching. Progressive enhancement of rather soft vibration modes should be observable, although they do not seem to play a particularly crucial role.

An interesting continuation of this work is to study this system with smaller and larger values of the contact friction coefficient $\mu$. We have seen that GG layers prepared with $\mu=0.1$ have a significantly smaller critical angle $\alpha_c \simeq 14^\circ$, whereas those prepared with $\mu=10$ ($\alpha_c \simeq 21^\circ$) do not differ much to the case $\mu=0.5$. Similarly, the grains are more connected and more densely packed at that critical angle in layers with $\mu=0.1$ ($Z\simeq 3.6$ and $\phi\simeq 0.833$), whereas the values of $Z$ and $\phi$ do not change much when the system is prepared with $\mu=10$. However, the systematic investigation of the role of $\mu$ on the mechanical response of the granular layer, and on the elastic shear modulus in particular, is beyond the scope of the present paper. Another possible perspective could be to use granular simulations with a rolling resistance \cite{ETR08} in order to explore a wider range of $\phi$, $Z$ and $\alpha$.

\acknowledgments
We thank B. Andreotti, E. Cl\'ement, J. Kurchan and M. Trulsson for usefull discussions and a careful reading of the manuscript. This work is part of the ANR JamVibe, project \# 0430 01. A.P.F. Atman has been partially supported by the exchange program `Science in Paris 2010' (Mairie de Paris). A.P.F. Atman thanks CNPq and FAPEMIG Brazilian agencies for financial funding.



\begin{thebibliography}{0}

\bibitem{vH10}
  \Name{van Hecke M.}
  \REVIEW{J. Phys. Cond. Mat.}{22}{2010}{033101}.

\bibitem{LNvSW10}
  \Name{Liu A.J., Nagel S.R., van Saarloos W. \and Wyart M.}
  \Book{Dynamical Heterogeneities in Glasses, Colloids, and Granular Media}
  \Editor{L. Berthier, G. Biroli, J.-P. Bouchaud, L. Cipelletti and W. van Saarloos}
  \Publ{Oxford Univ. Press, Oxford,UK}
  \Year{2011}
  \Page{298}.

\bibitem{oHLLN02}
  \Name{O'Hern C.S., Langer S.A., Liu A.J. \and Nagel S.R.}
  \REVIEW{Phys. Rev. Lett.}{88}{2002}{075507}.
  
\bibitem{oHSLN03}
  \Name{O'Hern C.S., Silbert L.E., Liu A.J. \and Nagel S.R.}
  \REVIEW{Phys. Rev. E}{68}{2003}{011306}.

\bibitem{MSLB07}
  \Name{Majmudar T.S., Sperl M., Luding S. \and Behringer R.P.}
  \REVIEW{Phys. Rev. Lett.}{98}{2007}{058001}.
  
\bibitem{WNW05}
  \Name{Wyart M., Nagel S.R. \and Witten T.A.}
  \REVIEW{Europhys. Lett.}{72}{2005}{486}.
  
\bibitem{EvHvS09}
  \Name{Ellenbroek W.G., van Hecke M. \and van Saarloos W.}
  \REVIEW{Phys. Rev. E}{80}{2009}{061307}.

\bibitem{R00}
  \Name{Roux J.-N.}
  \REVIEW{Phys. Rev. E}{61}{2000}{6802}.

\bibitem{M01}
  \Name{Moukarzel C.F.}
  \REVIEW{Granular Matter}{3}{2001}{41}.

\bibitem{TW99}
  \Name{Tkachenko A.V. \and Witten T.A.}
  \REVIEW{Phys. Rev. E}{60}{1999}{687}.

\bibitem{AR07}
  \Name{Agnolin I. \and Roux J.-N.}
  \REVIEW{Phys. Rev. E}{76}{2007}{061302},
  \REVIEW{Phys. Rev. E}{76}{2007}{061303},
  \REVIEW{Phys. Rev. E}{76}{2007}{061304}.

\bibitem{SLN05}
  \Name{Silbert L.E., Liu A.J. \and Nagel S.R.}
  \REVIEW{Phys. Rev. Lett.}{95}{2005}{098301}.

\bibitem{DHRR05}
  \Name{Drocco J.A., Hastings M.B., Reichhardt C.J.O. \and Reichhardt C.}
  \REVIEW{Phys. Rev. Lett.}{95}{2005}{088001}.

\bibitem{SBoHS11}
  \Name{Schreck C.F., Bertrand T., O'Hern C.S. \and Shattuck M.D.}
  \REVIEW{Phys. Rev. Lett.}{107}{2011}{078301}.

\bibitem{GCSKB10}
  \Name{Ghosh A., Chikkadi V.K., Schall P., Kurchan J. \and Bonn D.}
  \REVIEW{Phys. Rev. Lett.}{104}{2010}{248305}.

\bibitem{KGMI10}
  \Name{Kaya D., Green N.L., Maloney C.E. \and Islam M.F.}
  \REVIEW{Science}{329}{2010}{656}.

\bibitem{CEZCYHBDvSLY10}
  \Name{Chen K., Ellenbroek W.G., Zhang Z., Chen D.T.N., Yunker P.J., Henkes S., Brito C., Dauchot O., van Saarloos W., Liu A.J. \and Yodh A.G.}
  \REVIEW{Phys. Rev. Lett.}{105}{2010}{025501}.

\bibitem{SvHESvS07}
  \Name{Somfai E., van Hecke M., Ellenbroek W.G., Shundyak K. \and van Saarloos W.}
  \REVIEW{Phys. Rev. E}{75}{2007}{020301}.

\bibitem{HvHvS10}
  \Name{Henkes S., van Hecke M. \and van Saarloos W.}
  \REVIEW{Europhys. Lett.}{90}{2010}{14003}.

\bibitem{HSvSvH10}
  \Name{Henkes S., Shundyak K., van Saarloos W. \and van Hecke M.}
  \REVIEW{Soft Matter}{6}{2010}{2935}.

\bibitem{S10}
  \Name{Silbert L.E.}
  \REVIEW{Soft Matter}{6}{2010}{2918}.

\bibitem{BDBB10}
  \Name{Brito C., Dauchot O., Biroli G. \and Bouchaud J.-P.}
  \REVIEW{Soft Matter}{6}{2010}{3013}.

\bibitem{LN98}
  \Name{Liu A.J. \and Nagel S.R.}
  \REVIEW{Nature}{396}{1998}{21}.

\bibitem{LN10}
  \Name{Liu A.J. \and Nagel S.R.}
  \REVIEW{Soft Matter}{6}{2010}{2869}.

\bibitem{BZCB11}
  \Name{Bi D., Zhang J., Chakraborty B. \and Behringer R.P.}
  \REVIEW{Nature}{480}{2011}{355}.

\bibitem{DEM}
  \Name{Rapaport D.C.}
  \Book{The Art of Molecular Dynamics Simulation}
  \Publ{Cambridge Univ. Press, Cambridge,UK}
  \Year{1995}.
  
\bibitem{HB09}
  \Name{Heussinger C. \and Barrat J.-L.}
  \REVIEW{Phys. Rev. Lett.}{102}{2009}{218303}.
  
\bibitem{SVR02}
  \Name{Staron L., Vilotte J.-P. \and Radja\"\i F.}
  \REVIEW{Phys. Rev. Lett.}{89}{2002}{204302}.

\bibitem{HBDvS10}
  \Name{Henkes S., Brito C., Dauchot O. \and van Saarloos W.}
  \REVIEW{Soft Matter}{6}{2010}{2939}.

\bibitem{dCEPRC05}
  \Name{da Cruz F., Emam S., Prochnow M., Roux J.-N. \and Chevoir F.}
  \REVIEW{Phys. Rev. E}{72}{2005}{021309}.
  
\bibitem{OH11}
  \Name{Otsuki M. \and Hayakawa H.}
  \REVIEW{Phys. Rev. E}{83}{2011}{051301}.

\bibitem{ABGRCCBC05}
  \Name{Atman A.P.F., Brunet P., Geng J., Reydellet G., Combe G., Claudin P., Behringer R.P. \and Cl\'ement E.}
  \REVIEW{J. Phys. Cond. Mat.}{17}{2005}{S2391}.

\bibitem{GACCG06}
  \Name{Goldenberg C., Atman A.P.F., Claudin P., Combe G. \and Goldhirsch I.}
  \REVIEW{Phys. Rev. Lett.}{96}{2006}{168001}.

\bibitem{bookDEM}
  \Book{Discrete-element modeling of granular materials}
  \Editor{F. Radja\"\i\ \and F. Dubois}
  \Publ{ISTE, Wiley}
  \Year{2011}

\bibitem{SRCCL01}
  \Name{Serero D., Reydellet G., Claudin P., Cl\'ement E. \and Levine D.}
  \REVIEW{Eur. Phys. J. E}{6}{2001}{169}.

\bibitem{GG02}
  \Name{Goldhirsch I. \and Goldenberg C.}
  \REVIEW{Eur. Phys. J. E}{9}{2002}{245}.

\bibitem{LTWB04}
  \Name{Leonforte F.,Tanguy A., Wittmer J.P. \and Barrat J.-L.}
  \REVIEW{Phys. Rev. B}{70}{2004}{014203}.

\bibitem{GG05}
  \Name{Goldhirsch I. \and Goldenberg C.}
  \REVIEW{Nature}{435}{2005}{188}.
  
\bibitem{ABGRCBC05}
  \Name{Atman A.P.F., Brunet P., Geng J., Reydellet G., Claudin P., Behringer R.P. \and Cl\'ement E.}
  \REVIEW{Eur. Phys. J. E}{17}{2005}{93}.

\bibitem{ACCGG09}
  \Name{Atman A.P.F., Claudin P., Combe G., Goldenberg C. \and Goldhirsch I.}
  \Book{Proc. 6th Int. Conf. Micromechanics of Granular Media, Powders and Grains 2009}
  \Editor{M. Nakagawa and S. Luding}
  \Publ{American Inst. Physics}
  \Year{2009}
  \Page{492}
  
\bibitem{ACC09}
  \Name{Atman A.P.F., Claudin P. \and Combe G.}
  \REVIEW{Comput. Phys. Comm.}{180}{2009}{612}.

\bibitem{WSNW05}
  \Name{Wyart M., Silbert L.E., Nagel S.R. \and Witten T.A.}
  \REVIEW{Phys. Rev. E}{72}{2005}{051306}.

\bibitem{OBCS03}
  \Name{Otto M., Bouchaud J.-P., Claudin P. \and Socolar J.E.S.}
  \REVIEW{Phys. Rev. E}{67}{2003}{031302}.

\bibitem{BAC08}
  \Name{Bonneau L., Andreotti B. \and Cl\'ement E.}
  \REVIEW{Phys. Rev. Lett.}{101}{2008}{118001}.

\bibitem{ETR08}
  \Name{Estrada N., Taboada A. \and Radja\"\i\ F.}
  \REVIEW{Phys. Rev. E}{78}{2008}{021301}.


\end{thebibliography}
\end{document}